\documentclass[amsfonts,amsmath,prd,preprint,nofootinbib]{revtex4}

\usepackage{epsfig,bbm,cancel,ulem}
\usepackage[breaklinks=true]{hyperref}
\usepackage{xcolor}
\usepackage{wasysym}
\usepackage[mathscr]{eucal}
\usepackage{multirow}
\usepackage{amsmath}
\usepackage{enumitem}
\usepackage{array}
\usepackage{tabularx}
\usepackage{orcidlink}

\begin{document}

\title{Non-extensive entropy signatures in compact star}

\author{B. N. Jayawiguna \orcidlink{0000-0001-5920-8701}}
\email{nugrahabyon312@gmail.com}
\author{M. H. Al Ghifari \orcidlink{0009-0005-9545-4498}}
\email{muhammad.hafizt31@ui.ac.id}
\author{A. Rohim \orcidlink{0000-0002-6629-9979}}
\email{ar.rohim@ui.ac.id}
\author{A. Sulaksono \orcidlink{0000-0002-1493-5013}}
\email{anto.sulaksono@sci.ui.ac.id}
\affiliation{Departemen Fisika, FMIPA, Universitas Indonesia, Depok, 16424, Indonesia.}
\def\changenote#1{\footnote{\bf #1}}

\begin{abstract}
The non-extensive entropy models applied to the black-hole horizon are connected to the generalized Einstein--Hilbert action, $f(R)$, via the Wald entropy formalism. We demonstrate that quark star configurations characterized by the MIT bag model under this modified gravity approach conform to the observational lower and upper bounds established for compact objects linked to HESS J1731$-$347 and GW190814. Furthermore, we assess the effective energy conditions and corresponding speed of sound to evaluate the physical plausibility and stability of the stellar configurations. All  non-extensive entropy models studied satisfied these conditions.
Our findings show that the exotic compact objects offer a compelling and credible framework for exploring aspects of non-extensive entropy models, such as the Barrow, Tsallis-Cirto, and R\'enyi formalisms, and vice versa.
\end{abstract}

\maketitle
\thispagestyle{empty}

\setcounter{page}{1}

\section{Introduction}
Understanding spacetime's structure across all scales requires clarity in the interplay of gravity, thermodynamics, and quantum theory. Black holes serve as exact platforms for probing these connections. Bekenstein and Hawking, using a semi-classical approach, demonstrated that black holes behave as thermodynamic systems with temperature and entropy \cite{Bekenstein1973,Hawking1975}. Notably, classical thermodynamics ties entropy to volume, while Bekenstein-Hawking entropy is proportional to horizon area, revealing a non-extensive trait. The source of this non-extensiveness remains unresolved \cite{Nojiri2022}. Horizonless compact objects might shed light on the nature of this puzzle. Ongoing research targets quantum gravity effects to refine entropy calculations, using the Bekenstein-Hawking result as the baseline and accounting for further logarithmic, power-law, or exponential corrections \cite{Anand:2025rjg}. Recent advances have elevated interest in Bekenstein-Hawking entropy generalizations, including Tsallis, Barrow, Rényi, and Kaniadakis forms \cite{Tsallis:1987eu,Barrow:2020tzx,renyi,Kaniadakis:2005zk}. See \cite{Anand:2025rjg,Nojiri2022} and references therein for discussion of impacts on black holes and cosmology. Additional analyses of more generalized entropies for cosmological and black hole contexts appear in \cite{Odintsov2023,Luciano2026}. The recent observations from the entropy model have been widely discussed in the literature. The Tsallis entropy framework has been employed to modify the Friedmann equation in the context of dark matter phenomenology constrained by the \textit{IceCube} high-energy neutrino observations, yielding a characteristic non-extensive parameter of approximately $\delta \simeq 1.565$. Furthermore, Big-Bang nucleosynthesis (BBN) constraints derived from the relic abundance analysis of cold dark matter (CDM) particles indicate a preferred value of $\delta \simeq 1.499$ \cite{Jizba:2023fkp}. Related investigations within non-extensive cosmology and modified gravitational frameworks, including extensions based on Barrow entropy, have also been extensively explored in Refs.~\cite{AlMamon:2020usb,Luciano:2022hhy,Luciano:2022ely}.

General relativity (GR) can be extended by incorporating higher-order curvature invariants into the action, with $f(R)$ theories serving as prominent examples. Comprehensive reviews of $f(R)$ gravity theory are available in Refs.~\cite{Sotiriou2010,DeFelice2010,Nojiri2017}. The extensions in $f(R)$ theory are not constructed arbitrarily; rather, they can arise naturally within an effective field theory framework for gravity coupled to matter fields, consistent with the renormalization structure of quantum field theory in curved spacetime (see Ref. \cite{DAgostino:2025axy} and references therein for further details). It is worth noting that several functional forms of $f(R)$ have been proposed in the literature (see \cite{Numajiri:2023uif,Nashed:2021lzq,Cui:2024nkr,Chen:2024wqc}). These different $f(R)$ gravity theories can raise different issues, such as stability conditions and the compatibility of their predictions with observational data. Please see Refs. ~\cite{Santos:2007bs,Pretel:2022plg} and the references therein for detail. Particularly, for horizonless compact objects such as white dwarfs, neutron stars, and quark stars, like other modified gravity theories, these theories modify the stellar structures, such as mass-radius relation, moment of inertia, and tidal deformation predictions, as well as the stability of the model. Please see Refs. \cite{Olmo2020,Danarianto2025} for details. 

A direct connection exists between entropy evaluated at the horizon and $f(R)$ theory through the Wald entropy formulation \cite{wald,Iyer:1994ys}. The relationship between $f(R)$ theory and Tsallis entropy via Wald entropy is discussed in Ref. \cite{DAgostino2024}, while the association between the generalized uncertainty principle entropy and $f(R)$ theory is addressed in Refs. \cite{DAgostino:2025axy,Hammad2015,Nojiri:2021czz,Odintsov:2022qnn,Nojiri:2023wzz,DAgostino:2024sgm}. These connections provide a promising framework for testing corrections to GR through astrophysical and cosmological observations. To date, such tests have primarily focused on black holes and cosmological phenomena, whereas investigations of the non-extensive Bekenstein-Hawking entropy generalization in horizonless compact objects remain limited. In this work, we try to fulfill this gap by further exploring horizonless compact objects, especially quark stars, as candidates for probing these theoretical advancements.

Quark stars are interesting because they represent a theoretical "third state" of stellar death—denser than neutron stars but not quite black holes—providing a unique laboratory to test fundamental physics at extreme densities. They are hypothesized to be composed of "strange quark matter," a deconfined soup of quarks that could be the universe's most stable form of matter \cite{Kumari2021}. Please see the recent review of quark stars and the corresponding equation of state (EoS) in Refs.~\cite{Danarianto2025,Zhang2024} for details. The anisotropic quark stars in $f(R) = R^{1+\epsilon}$ gravity with $\epsilon \ll 1$ using MIT Bag model EoS are studied by Pretel $\it et ~ al$~\cite{Pretel:2022plg}. Unlike in the GR case, the $f(R)$ influences the exterior solution, and hence, there must exist a numerical technique to match between the interior and exterior at the surface, and also match with the pure Schwarzschild solution in the asymptotic region. In addition, the quark star's mass radius depends on the anisotropic strength and the small parameter $\epsilon$. They found that the small Ricci scalar correction, $\epsilon$, has significantly changed the mass and radius. With the fixed $\epsilon$ value, the maximum mass increases (decreases) when the anisotropic parameter is positive (negative). Moreover, in the recent observations, the nature of the massive secondary objects detected by LIGO and Virgo in the GW190814 event, with masses of 2.50–-2.67 $M_\odot$, remains mysterious as their physical properties do not clearly categorize them as neutron stars or black holes \cite{LIGOScientific:2020zkf}. Discussions about the nature of this object are ongoing, addressing whether it is a light black hole, a fast-rotating neutron star, a quark star, or a hybrid star; there is no consensus \cite{Annala2020}. Please also see Ref.\cite{Prasetyo2021} for more details. HESS J1731-347~\cite{Doroshenko:2022nwp,Horvath:2023uwl} is another exotic compact object, a shell-type supernova remnant containing an unusually light and compact, estimated at $\sim 1M_{\odot }$ and a radius of $\sim 10.4$ km. It is possibly the lightest neutron star detected or could represent a rare exotic compact object, such as a quark star, a hybrid star, or a dark-matter-admixed neutron star \cite{Bhat2020}. If both GW190814's secondary object and HESS J1731-347 are assumed to be quark stars, this opens the possibility to examine signatures of Bekenstein Hawking entropy generalizations, including Tsallis, Barrow, Rényi, and Kaniadakis forms \cite{Tsallis:1987eu, Barrow:2020tzx, renyi, Kaniadakis:2005zk}. 

In the present work, we investigate the Wald entropy formalism to establish a connection between non-extensive entropy models and $f(R)$ gravity. To further analyze the modified field equations, we employ the MIT bag model equation of state (EoS) within quark star configurations and examine the corresponding mass--radius relations under observational constraints from the low-mass regime, represented by HESS J1731-347, and the massive regime associated with GW190814. Furthermore, in contrast to the standard General Relativity (GR) framework, $f(R)$ gravity modifies the energy conditions into effective forms, leading to characteristic deviations from their GR counterparts. We therefore compare the modified gravity predictions with the corresponding GR results. Throughout this work, we adopt geometrized units with $G=c=1$. Therefore, all physical quantities are expressed in terms of length.

This paper is organized as follows. In Section II, we talk about the gravity action and the changes to the Einstein field equations in $f(R)$ gravity for a simple perfect fluid inside, the field equation outside, and the rules for matching to the Schwarzschild solution far away. Section III explains the link between non-extensive entropy models and $f(R)$ gravity using the Wald entropy method. Section IV applies this idea to several well-known non-extensive entropy models in the literature: Tsallis--Cirto, Barrow, Rényi, and Kaniadakis. Section V uses the MIT bag model EoS within the modified field equations to calculate mass-radius relations, effective energy relations, and sound-speed profiles in quark stars at different central pressures. Section VI contains the conclusions.

\section{Entropy in GR}

The discovery of Hawking radiation established a profound connection between gravity and thermodynamics, demonstrating that the four laws of black hole mechanics are closely analogous to the laws of thermodynamics. Furthermore, it was later shown that these thermodynamic laws can be derived from the Einstein field equations, suggesting that gravitation possesses an intrinsic thermodynamic character. This naturally raises the inverse question: can the Einstein field equations themselves be derived from thermodynamic principles? In a seminal work, Jacobson showed in 1995 that the Einstein field equations can indeed be obtained by using the first law of local Rindler horizons, thereby interpreting the field equations as an equation of state for spacetime \cite{Jacobson:1995ab}. Motivated by this thermodynamic perspective, Gao \cite{Gao:2011hh} employed the maximum entropy principle to derive the (TOV) equation for self-gravitating systems.

In this section, we present a pedagogical introduction to the role of entropy in general relativity and subsequently extend the discussion to modified theories of gravity. We begin with the thermodynamics of black holes, where the notion of gravitational entropy was first established, and then proceed to self-gravitating stellar configurations. The Einstein--Hilbert action of general relativity is given by
\begin{equation}
\mathcal{A} = \int d^{4}x  \sqrt{-g}~ R + \int d^{4}x  \sqrt{-g}  \mathcal{L}_{m},
\end{equation}
where the $g$ is the determinant of the metric tensor, $R$ is the Ricci scalar, and the second term contains the matter Lagrangian. The variation of the metric tensor will give us the Einstein field equation
\begin{equation}
R_{\mu\nu} - \frac{1}{2}g_{\mu\nu}R = 8\pi T_{\mu\nu}.
\end{equation}
In the vacuum case, the tensor energy-momentum $T_{\mu\nu}=0$, and the solution can be considered as Schwarzschild solution
\begin{equation}
ds^2 = -\left(1- \frac{2M}{r}\right) dt^2 + \left(  1-\frac{2M}{r} \right)^{-1} dr^2 + r^2 d\theta^2 + r^2 \sin^2\theta d\phi^2,
\end{equation}
where $M$ is the total mass of the object. Note that the solution above is the exterior solution of the object with mass $M$.

\subsection{BH's Entropy}

In the BH case \cite{Bardeen:1973gs}, it has been reported that the semi-classical black hole properties obey a certain law of thermodynamics
\begin{equation}
dM = T dS + \Omega dJ,
\end{equation}
where $T$, $S$,~$\Omega$,~and $J$ are temperature, entropy, angular velocity, and angular momentum, respectively. In the static case ($J=0$), the entropy can be expressed as
\begin{equation}
S = \int \frac{1}{T} \frac{\partial M}{\partial r_{+}} dr_{+}.
\end{equation}
From here, it is shown that the thermodynamic quantities can be extracted from the outer horizon BH with the condition $f(r_{+})=0$. Combining with the Ricci scalar at the horizon, $R_{+}=2/r_{+}^2$, the entropy expression reads
\begin{equation}
	\label{sgr}
S = \frac{2\pi}{R_{+}}.
\end{equation}
In the particular Schwarzschild case, the Hawking temperature and entropy can be written as
\begin{eqnarray}
T_{H}= \frac{1}{4\pi r_{+}},~~~~\textrm{and}~~~~ S = \pi r_{+}^2.
\end{eqnarray}
The Bekenstein entropy is nonextensive, and it obeys the following nonadditive
composition rule for the sum of the entropies of two (spatial) disjoint horizons $S_{A}$ and $S_{B}$ \cite{Alonso-Serrano:2020hpb}.
\begin{equation}
S_{A+B} = S_{A} + S_{B} + 2\sqrt{S_{A}} \sqrt{S_{B}}. 
\end{equation}
 Please also see the discussion in Ref. \cite{Dabrowski:2024qkp} for the Tsallis entropy case.

\subsection{Star's Entropy}

In the star's case, the properties are merely different since it is a horizonless object and matter plays a significant role in the thermodynamics of this object. In this case, the corresponding discussion has already been performed in detail in Refs. \cite{Gao:2011hh} and \cite{Sorkin:1981wd}, here, we will summarize what they had found. For the general perfect fluid case, we need a different prescription from the BH case; namely, we start from the Boltzmann-Gibbs entropy density expression, $s$, and treat $(\rho,n)$ as two independent variables. They start with the unit-volume version of the first law of standard thermodynamics as
\begin{equation}
	ds = \frac{1}{T} d\rho - \frac{\mu}{T} dn,
\end{equation}	
where $s=S/V$, and $n=N/V$. The total entropy and particle number read
\begin{eqnarray}
	S &=& 4\pi \int_{0}^{R} s(r) \left[1-\frac{2m(r)}{r}\right]^{-1/2} r^2 dr, \\ N &=& 4\pi \int_{0}^{R} n(r) \left[1-\frac{2m(r)}{r}\right]^{-1/2} r^2 dr.
\end{eqnarray}
In this case, the geometry (metric) enters into entropy through the Lorentz contraction factor 
$\left[1-\frac{2m(r)}{r}\right]^{-1/2}$. By applying the maximum entropy principle and augmented with the Lagrange multiplier method, the results are 
\begin{eqnarray}
	\mu &=& \lambda T,\\ \label{eq2}
	(4\pi p r^3 +m) T +(r-2m)rT'&=&0,
\end{eqnarray}
respectively. Using the Gibbs-Durhem relation and the definition $T=T(\rho,\mu)$, we get
\begin{equation}
	\label{Tprime}
	T'=\frac{T}{\rho+p} p'(r).
\end{equation}
Hence, by inserting \eqref{Tprime} into \eqref{eq2}, they found the general expression for the TOV equation that can also be obtained from the Einstein field equations as
\begin{equation}
	\frac{dp}{dr} = -\frac{(\rho+p)[m(r)+4\pi r^3 p]}{r[r-2m(r)]}.
\end{equation}

Note that the derivation of the TOV equation based on thermodynamics has also been extended to charged perfect fluid configurations \cite{Gao:2011hh}. In conclusion, these results demonstrate that the TOV equation, which was originally derived from the gravitational field equations and the conservation of the perfect fluid energy--momentum tensor, can equivalently be obtained from the standard laws of thermodynamics through the maximum entropy principle. This equivalence provides good evidence for the deep and fundamental connection between gravitation and thermodynamics in horizonless objects.

It is important to emphasize, however, that the notion of entropy differs significantly between black holes and horizonless compact objects. For black holes within GR, the entropy is a geometric quantity determined entirely by the properties of the event horizon, being proportional to its area according to the Bekenstein--Hawking relation. In contrast, the entropy of a horizonless compact object originates from the thermodynamic degrees of freedom of the matter composing the star. It is obtained by integrating the local entropy density throughout the stellar interior. Consequently, the stellar entropy is an extensive thermodynamic quantity, scaling with the amount of matter in the system. By contrast, black hole entropy is intrinsically non-extensive, whereby the entropy scales with the horizon area rather than the system volume. This fundamental distinction underlies the different thermodynamic descriptions of black holes and self-gravitating stars.

\section{Entropy principle in $f(R)$ gravity}

 Now, we shift the attention of the previous case to the more general case, $f(R)$ gravity \cite{Fang:2015pcw}. We show both BH and star's entropy in two different subsection to make it more pedagogical.

\subsection{BH's Entropy}

It is well known that the standard Bekenstein-Hawking entropy~\cite{Bekenstein1973,Hawking1975} is formulated within the classical GR theory. However, in the generalized gravity theories such as including quantum effect, the direct relationship between entropy and area is not straightforward. It is both mathematically and physically expected, given the curvature term in the theory. Therefore, we prefer to consider the general entropy expression coming from the Wald formulation \cite{wald,Iyer:1994ys}
\begin{equation}
	\label{WE}
	S_{\textrm{Wald}} = \frac{1}{4} \int_{\mathcal{H}} d^{2}x \sqrt{h} f_{R}(R),
\end{equation}
where the quantity $\mathcal{H}$ denotes the BH horizon cross section, $h$ is the induced metric on $\mathcal{H}$, and $f_R (R)= df(R)/dR.$ Moreover, the Ricci scalar  in the spherically symmetric spacetime evaluated at the horizon $R_{+}$ with radius $r_{+}$ is given by
\begin{equation}
	\label{Rr}
	R_{+} = \frac{2}{r_{+}^2}~~~~ \rightarrow~~~~r_{+}=\sqrt{\frac{2}{R_{+}}}.
\end{equation}
Thus, the black hole entropy evaluated at the horizon $S(r_+)$ can be expressed as $S(R_+)$. Furthermore, using Eq.~\eqref{Rr}, the integral of Wald entropy in Eq. \eqref{WE} can be simplified into \cite{Zheng:2018fyn}
\begin{equation}
	S_{\textrm{Wald}} = \frac{2\pi}{R_+} f_{R}(R_+),
\end{equation}
where it is in agreement with results obtained by using the Euclidean
semiclassical approach or the Wald entropy formula \cite{Dyer:2008hb,Vollick:2007fh,Iyer:1995kg}. The same case can also be applied to the $f(R,R_{\mu\nu}R^{\mu\nu})$ gravity \cite{Feng:2019ejb}. In GR case ($f_{R}=1$), the entropy expression reduces to eq \eqref{sgr}. In Ref. \cite{DAgostino:2025axy}, they identified the Wald entropy expression as the GUP entropy. Here, using the same reasoning, we identified the non-extensive entropies ${S_i}$ as a generalized expression of Bekenstein-Hawking entropy as Wald (non-extensive) entropy. In this way, we can find the $f_R$ relation in terms of the Ricci scalar evaluated at the horizon of the black hole. Note that $i$ is model indexed. The generalized $f(R_+)$ can be obtained by integrating with respect to $R_+$
	\begin{equation}
		\label{fR}
		f(R_+) \equiv \int \frac{S_i R_+}{2\pi} dR_+ + c_1,
	\end{equation}
	where $c_1$ is the constant integration that can be interpreted as the cosmological constant \cite{DAgostino:2025axy}. But for now, we neglect the cosmological constant since the values are small \cite{Singh:2024bcd} . Similar to GR, we can assume that the form of the $f(R)$ theory in stars and black holes is the same. Therefore, we can omit the "+" index in the $f(R)$ form obtained from Eq. \eqref{fR} for star applications. In this way, we can consider this relation as a connection between the compact star evaluated by the general $f(R)$ and the non-extensive entropy. Furthermore, because $S_\textrm{ Wald}$ merely depends on the horizon area through $f_R (R_+)$, we can claim that all extensions of Bekenstein-Hawking entropy through Wald entropy are also non-extensive.

\subsection{Star's Entropy}

Having established the entropy of black holes, we now turn to the entropy of self-gravitating stars, which is expected to differ from the Bekenstein--Hawking prescription due to the absence of an event horizon. Following Ref.~\cite{Fang:2015pcw}, we present two complementary derivations of the TOV equation. The first follows the conventional approach, deriving the TOV equation from the gravitational field equations together with the conservation of the energy--momentum tensor. The second adopts a purely thermodynamic perspective, in which the TOV equation is obtained by applying the maximum entropy principle to a self-gravitating fluid configuration.

\subsubsection{Standard Method}

In this section, we briefly discuss the Einstein field equations using the $f(R)$ theory coupled with the perfect fluid matter sector for both interior and exterior evaluation and boundary conditions. The action of the $f(R)$ gravity can be written as

\begin{equation}
	\mathcal{A}=\frac{1}{16} \int d^4 x \sqrt{-g} f(R) + S_{\textrm{m}},
\end{equation}
where the $g$ is the determinant of the metric tensor $g_{\mu\nu},$ $R$ is the Ricci scalar, and the $S_{\textrm{m}}$ is a quantity related to the action matter. By varying the above equation with the contravariant metric tensor, the final field equation is
\begin{equation}
	f_{R}R_{\mu\nu}-\frac{1}{2}g_{\mu\nu}f-\nabla\mu\nabla\nu f_{R}+ g_{\mu\nu} \Box f_{R} = 8\pi T_{\mu\nu},
\end{equation}
where the $\nabla_{\mu}$, $\Box\equiv \nabla_{\mu}\nabla^{\nu}$, $f_{R}\equiv df/dR$, and $T_{\mu\nu}$ are the covariant derivative, d'Alembert operator, derivative with respect to $R$, and the tensor energy-momentum, respectively. Next, the interior of the stellar configurations can be assumed to be an isotropic perfect fluid such as
\begin{equation}
	T_{\mu\nu} = (\rho+p)u_{\mu}u_{\nu}+p g_{\mu\nu},
\end{equation}
with $u^{\mu}$ denotes the 4-velocity in which satisfies the normalization condition,$u_{\mu}u^{\mu}=-1.$ Furthermore, the $\rho$ is the energy density, and the $p$ is the pressure of the star. The trace of the tensor energy-momentum is given by $T = -\rho+3p$.

Having all of this in hand, we can evaluate the star's properties by defining the spherically symmetric ansatz for the metric tensor
\begin{equation}
	ds^2 = g_{\mu\nu}dx^{\mu}dx^{\nu} = -e^{\nu} dt^2 + e^{\lambda} dr^2 + r^2 (d\theta^2 + \sin\theta d\phi^2),
\end{equation}
where the metric tensor, in $tt$ and $rr$-component, depends only on $r.$ Therefore, we can write the radial components of the conservation law and d'Alembert operator to be

\begin{eqnarray}
	\label{tov}
	\frac{dp}{dr}=-\frac{(\rho+p)}{2}\nu',~~\textrm{and}~~ \Box f_{R} &=& \frac{1}{\sqrt{-g}} \partial_{\mu} \left[ \sqrt{-g} \partial^{\mu} f_{R}\right], \\ &=& e^{-\lambda} \left[ \left ( \frac{2}{r} + \frac{\nu'-\lambda'}{2} \right) f_{R}'+ f_{R}''\right],
\end{eqnarray}
where the prime denotes the derivative with respect to $r$. With what we have, we can move further by evaluating the Einstein field equations, each component, along with the aid of \eqref{tov}, to have
\begin{eqnarray}
	\label{nu}
	\frac{d}{dr}\bigg(\frac{\nu}{2} \bigg) &=&    \frac{r^2e^{\lambda}(16\pi p+f-Rf_{R})  +2f_{R} (e^{\lambda}-1) -4r R' f_{RR}}{2r(2f_{R}+rR'f_{RR})}  ,  \\ \frac{d}{dr}\bigg(\frac{\lambda}{2} \bigg) &=&   \frac{1}{2r(2f_{R}+rR' f_{RR})} \bigg \lbrace 2f_{R} (1-e^{\lambda}) +\frac{r^2 e^{\lambda}}{3} [16\pi(2\rho+3p)+Rf_{R}+f]    \nonumber \\ && +\frac{rR' f_{RR}}{f_{R}} \left[ 2f_{R} (1-e^{\lambda}) +\frac{r^2 e^{\lambda}}{3} (16\pi \rho+2Rf_{R}-f)+2rR'f_{RR}    \right]  \bigg \rbrace       ,   \\ \frac{d^2 R}{dr} &=&  \frac{1}{3f_{RR}} \bigg\lbrace e^{\lambda} [8\pi(-\rho+3p) +2f-Rf_{R}] -3R'^2 f_{RRR} \bigg\rbrace +\left( \frac{\nu'-\lambda'}{2}-\frac{2}{r} \right) R',
\end{eqnarray}
where the equations above describe the isotropic field equations in the $f(R)$ gravity \cite{Pretel:2022plg}. With the well-known $f(R)$ form, we can simultaneously solve these coupled equations, with the EoS definition $\rho\equiv\rho(p)$, along with the imposed boundary conditions in the near center as
\begin{eqnarray}
	p(0)=p_{c},~~~~\nu(0)=\nu_c,~~~~\lambda(0)=0,~~~~R(0)=R_{c},~~~~\textrm{and}~~~~ R'(0)=0,
\end{eqnarray}
since we have five variables to be determined. The surface of the star can be obtained by the condition,~$p(r_s)=0$. However, the star profile in the outside region is simpler since we no longer need the pressure and the density ($\rho=p=0$). The equations reduce to
\begin{eqnarray}
	\frac{d}{dr}\bigg(\frac{\nu}{2} \bigg) &=&    \frac{r^2e^{\lambda}(f-Rf_{R})  +2f_{R} (e^{\lambda}-1) -4r R' f_{RR}}{2r(2f_{R}+rR'f_{RR})}  ,  \\ \frac{d}{dr}\bigg(\frac{\lambda}{2} \bigg) &=&   \frac{1}{2r(2f_{R}+rR' f_{RR})} \bigg \lbrace 2f_{R} (1-e^{\lambda}) +\frac{r^2 e^{\lambda}}{3} [Rf_{R}+f]    \nonumber \\ && +\frac{rR' f_{RR}}{f_{R}} \left[ 2f_{R} (1-e^{\lambda}) +\frac{r^2 e^{\lambda}}{3} (2Rf_{R}-f)+2rR'f_{RR}    \right]  \bigg \rbrace       ,   \\ \frac{d^2 R}{dr} &=&  \frac{1}{3f_{RR}} \bigg\lbrace e^{\lambda} [2f-Rf_{R}] -3R'^2 f_{RRR} \bigg\rbrace +\left( \frac{\nu'-\lambda'}{2}-\frac{2}{r} \right) R',
\end{eqnarray}
and the EoS is no longer used in this region. Moreover, to get the complete solution, we have to make sure that the solution covers both interior and exterior by mathcing the solution with junction condition at the star surface
\begin{eqnarray}
	\label{conditions}
	[\nu]=0,~~~[\lambda]=0,~~~[R]=0,~~~[R']=0,
\end{eqnarray}
Here, $[x]$ represents the difference between the value of $x$ evaluated inside and outside at $\Sigma_{0}$, specifically $[x]=x^{-}|_{\Sigma_{0}}-x^{+}|_{\Sigma_{0}}$. In this convention, the symbols $+$ and $-$ denote quantities outside and inside the star, respectively, where $\Sigma_0=r_s$. It is important to note that the influence of $f(R)$ is not confined to the surface; consequently, its imprint persists. The exact Schwarzschild condition is recovered by choosing $f(R)=R$. In the generalized context, the Schwarzschild condition arises from the requirement of asymptotic flatness.
\begin{eqnarray}
	\lim_{r\rightarrow \infty} R(r)=0,~~~~  \lim_{r\rightarrow \infty} m(r)=\textrm{constant},~~~~\lim_{r\rightarrow \infty} \nu(r)=0
\end{eqnarray}
means that the $R_{c}$, and $\nu_c$ must be chosen so that the requirement above is satisfied at infinity.

From this, we can infer that the complete solution depends on the specific  $f(R)$ model. Inspired by \cite{DAgostino:2025axy}, we construct the $f(R)$ model based on several representative non-extensive entropy models.

\subsubsection{Thermodynamical Method}
In the compact object, we generally obtain the TOV equation by using the conservation equation in the radial coordinate and the Einstein field equation. The final result is presented in the eq \eqref{tov} combined with \eqref{nu}. The thermodynamical method can also be proven by using the Einstein field equation and the extrema of the total entropy. The procedure is similar with the GR case, the discrepancy starts in the constrained Euler-Lagrange. Using the same condition in \eqref{Tprime}, they obtain the equation \cite{Fang:2015pcw}
\begin{eqnarray}
	\label{pp}
	\frac{p' r^2}{\rho+p} \left( \frac{2}{r^2} f_{R}(R) + \frac{1}{r} f'  \right)  \left(1-\frac{2m}{r}\right) &=& \left(-\frac{4}{r^3} f_{R} + \frac{1}{r^2} f_{R}' + \frac{1}{r} f_{R}''   \right) r^2 \left(1-\frac{2m}{r}\right) \nonumber \\ && + \left( \frac{2}{r^2} f_{R} + \frac{1}{r} f_{R}' \right) (2r + rm' -5m) \nonumber \\ && - \left( \frac{2}{r} f_{R}''  + \frac{3}{r^2} f_{R}' \right) r^2 \left(1-\frac{2m}{r}\right) \nonumber \\ && -8\pi r(\rho+p).
\end{eqnarray} 
By substituting $tt$-component of the Einstein field equation for $f(R)$ gravity to eq \eqref{pp}, we obtain
\begin{eqnarray}
	-\frac{p'}{\rho+p} \left( \frac{2}{r^2} f_{R} + \frac{1}{r} f'  \right) r^2 \left(1- \frac{2m}{r}   \right) &=& 8\pi p + \frac{f}{2} + \left(\frac{2m}{r^3} -\frac{R}{2} \right) f_{R} - \frac{2(r-2m)}{r^2} f_{R}'.
\end{eqnarray}
The latter expression is the TOV equation of $f(R)$ gravity derived from the thermodynamical method, and again, the relation between gravity and thermodynamcs for horizonless objects holds in $f (R)$ theory.

\section{Some non-extensive entropies}
We briefly discuss several non-extensive entropies that have attracted significant attention. These entropies generalize standard BH thermodynamics, i.e., Bekenstein-Hawking entropy~\cite{Bekenstein1973,Hawking1975}.

\subsection{Tsallis-Cirto Entropy}
Tsallis entropy first appears in the study of non-extensive statistics for systems with long-range interactions, where the partition function diverges, and the standard Boltzmann-Gibbs entropy becomes inadequate \cite{Nojiri2022}. Building on this development, the standard Bekenstein-Hawking entropy receives a correction in the form of a power law. This generalized entropy, as a result, can accommodate the modified cosmological field equation, offering an alternative explanation for late-time cosmic acceleration. Specifically, the model \cite{Tsallis:1987eu} of the Tsallis-Cirto entropy is given by
\begin{equation}
	S_{\textrm{TS}} = (\pi r_+^2)^{1+\delta},
\end{equation}
where the quantity $\delta$ denotes the deformation parameter. The Bekenstein Hawking entropy $S_{BH}$ is recovered when $\delta=0$. 

Note that we slightly modify the form to get the same series expansion around $\delta\ll 1.$ First, the full expression obtained from Eq. \eqref{fR} reads
\begin{equation}
	f_{\textrm{TS}}(R) = \frac{(2\pi)^{\delta}}{1-\delta} R^{1-\delta},
\end{equation}
and its series expansion of $f(R)$ theory of Tsallis-Cirto entropy becomes 
\begin{equation}
	f_{\textrm{TS}}(R)\approx R + R\left(  1+\log \bigg| \frac{2\pi}{R}\bigg| \right) \delta+\mathcal{O}(\delta^2).
\end{equation}

\subsection{Barrow Entropy}
Formally, the Barrow entropy resembles Tsallis entropy. However, the physical principles underlying the two entropies differ. The Barrow entropy was proposed as a toy model for possible effects of gravitational spacetime foam. Parameter $\Delta$ quantifies the quantum gravitational deformation \cite{Nojiri2022}. The entropy model \cite{Barrow:2020tzx} can be written as
\begin{equation}
	S_{\textrm{B}} = (\pi r_+^2)^{1+\frac{\Delta}{2}}.
\end{equation}
Notice that the expression for Barrow entropy is similar to the Tsallis-Cirto one. The range of the parameter of the Barrow entropy is $0\leq\Delta \leq 1,$ while Tsallis-Cirto is $-1\leq\delta\leq1/2.$ At $\Delta=1$ (and $\delta=1/2$), the qualitative behaviour of the temperature as a function of the black hole mass is the same as the Tsallis-Cirto model. In this limit, both models yield an extensive, but still nonadditive entropy for black holes. The Bekenstein Hawking entropy case is recovered when we switch off the $\Delta.$ The $f_R$ form of this entropy becomes
\begin{equation}
	f_{R} = \left( \frac{2\pi}{R} \right)^{\Delta/2},
\end{equation}
where we used the relation \eqref{Rr}. The generalized Ricci scalar now reads
\begin{equation}
	f_{\textrm{B}}(R)= \frac{(2\pi)^{\Delta/2}}{\left(1-\frac{\Delta}{2}\right)} R^{1-\frac{\Delta}{2}} + c,
\end{equation}
where the constant $c$ can be considered as the cosmological constant \cite{DAgostino:2025axy}. In this time, we restrict the work without the $c$-term. In the very small parameter, $\Delta \ll 1,$ the form can be expanded into
\begin{equation}
	f_{\textrm{B}}(R) \approx R + \frac{\Delta R}{2} \left(1+\log\bigg| \frac{2\pi}{R} \bigg| \right) + \mathcal{O}(\Delta^2).
\end{equation}
The Ricci scalar in GR is located in the first term.

\subsection{Renyi Entropy}

Rényi entropy is a generalized additive entropy formalism extending the Boltzmann–Gibbs and Shannon framework through a deformation parameter that characterizes nonstandard probability distributions. The theory is widely used in information theory, multifractal systems, quantum entanglement, and gravitational thermodynamics. In astrophysics and black hole physics, Rényi entropy is applied to generalized horizon thermodynamics, black hole phase transitions, cosmological dynamics, and modified entropy-area relations in systems with long-range gravitational interactions. The Rényi entropy was proposed as an index of information and originally had no relation to the statistics of physical systems \cite{Nojiri2022}. The entropy can be written as \cite{renyi}
\begin{equation}
	S_{\textrm{R}} = \pi r_+^2 \frac{\log|1+\bar{\lambda}|}{\bar{\lambda}},
\end{equation}
where the $\bar{\lambda}=\lambda/\pi r_+^2,$ is the dimensionless quantity, and this expression can be seen as the $S_{BH}$ times the constant factor. As a result, the parameter $\bar{\lambda}$ determines the deviation from the classical GR. The $f(R)$ expression is slightly similar to the GR one.
\begin{equation}
	f_{\textrm{R}}(R)=R\frac{\log|1+\bar{\lambda}|}{\bar{\lambda}}.
\end{equation}

\subsection{Kaniadakis Entropy}

Kaniadakis entropy is a relativistic generalized statistical mechanics formalism based on $\kappa$-deformed logarithmic and exponential functions. The theory is motivated by relativistic kinetic theory and power-law statistical behavior in complex systems. In black hole physics and cosmology, Kaniadakis entropy is used to study generalized horizon thermodynamics, Hawking temperature corrections, black hole stability, and modified gravity. Among the Kaniadakis model, we choose the relevant model that has been used in the literature \cite{Anand:2025rjg}. The Kaniadakis entropy \cite{Kaniadakis:2005zk} model reads
\begin{equation}
	S_{\textrm{K}} = \frac{\sinh(\kappa~ S_{\textrm{BH}})}{\kappa},
\end{equation}
where the $\kappa$ is the constant for controlling the modified entropy. The Kaniadakis entropy reproduces the Bekenstein-Hawking entropy in the limit $\kappa \rightarrow 0$. This entropy can be considered as a generalization of the Boltzmann-Gibbs entropy arising in relativistic statistical systems\cite{Nojiri2022}. The $f(R)$ expression for Kaniadakis entropy reads

\begin{equation}
	f_{\textrm{K}}(R) = \frac{1}{2 \pi  \kappa } \left[ \frac{1}{2} R^2 \sinh \left(\frac{2 \pi \kappa }{R}\right)+\pi  \kappa 	R \cosh \left(\frac{2 \pi  \kappa}{R}\right)-2 \pi ^2 \kappa ^2 \text{Shi}\left(\frac{2 \pi \kappa}{R}\right) \right],
\end{equation}
where Shi(z) is denoted as the hyperbolic sine integral, $\int_{0}^{z} \text{sinh(t)}/\text{t} ~\text{dt}.$ As expected, in the limit of the small parameter, the GR expression is recovered 
\begin{equation}
	f_{\textrm{K}}(R) \approx R-\frac{2\pi^2\kappa^2}{3R} + \mathcal{O}(\kappa^3).
\end{equation}

In the next section, we examine four specific entropy models referenced above (in detail, see \cite{Anand:2025rjg}) to derive the associated generalized $f(R)$ gravity theory, and we study the quark stars (QS) within $f(R)$ obtained from these entropies, using the well-known MIT bag model for the equation of state (EoS) of quark matter. We also investigate whether the low mass and radius exotic compact star HESS J1731-347 and the high mass and radius compact star from GW 190814 are quark stars or not within these entropy models.
\section{Result and discussions}

In this section, we present the numerical results and the characteristic profile by plotting the mass radius and its energy conditions. In evaluating the field equations, one needs to specify the EoS, namely $\rho\equiv\rho(p)$. The QS is explored within $f(R)$ gravity using the well-known MIT bag model.
\begin{equation}
	\rho =3p+4B.
\end{equation}
The latter describes a fluid composed of down, up, and strange quarks. The $B$ denotes a bag constant with the value $B=60 $ MeV/fm$^3$ \cite{Paschalidis:2016vmz}.
\begin{figure}[h!]
	\centering
	\includegraphics[width=0.45\textwidth]{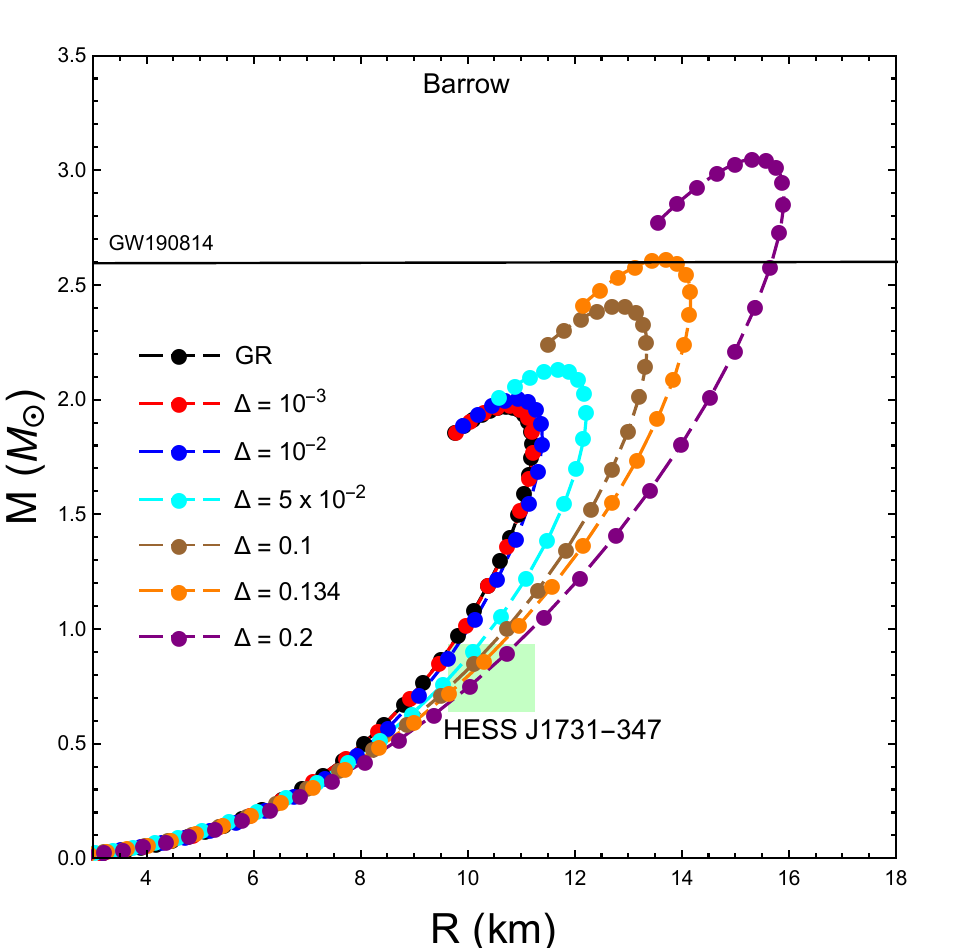}
	\includegraphics[width=0.45\textwidth]{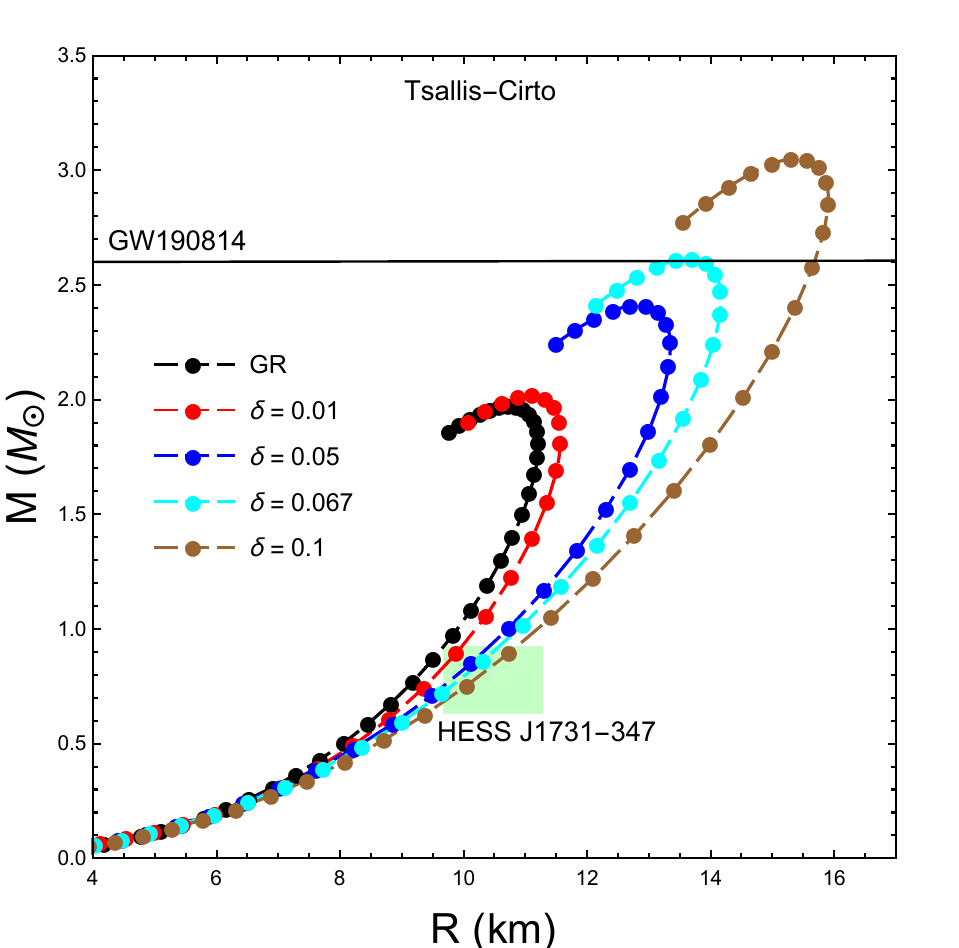}
	\includegraphics[width=0.45\textwidth]{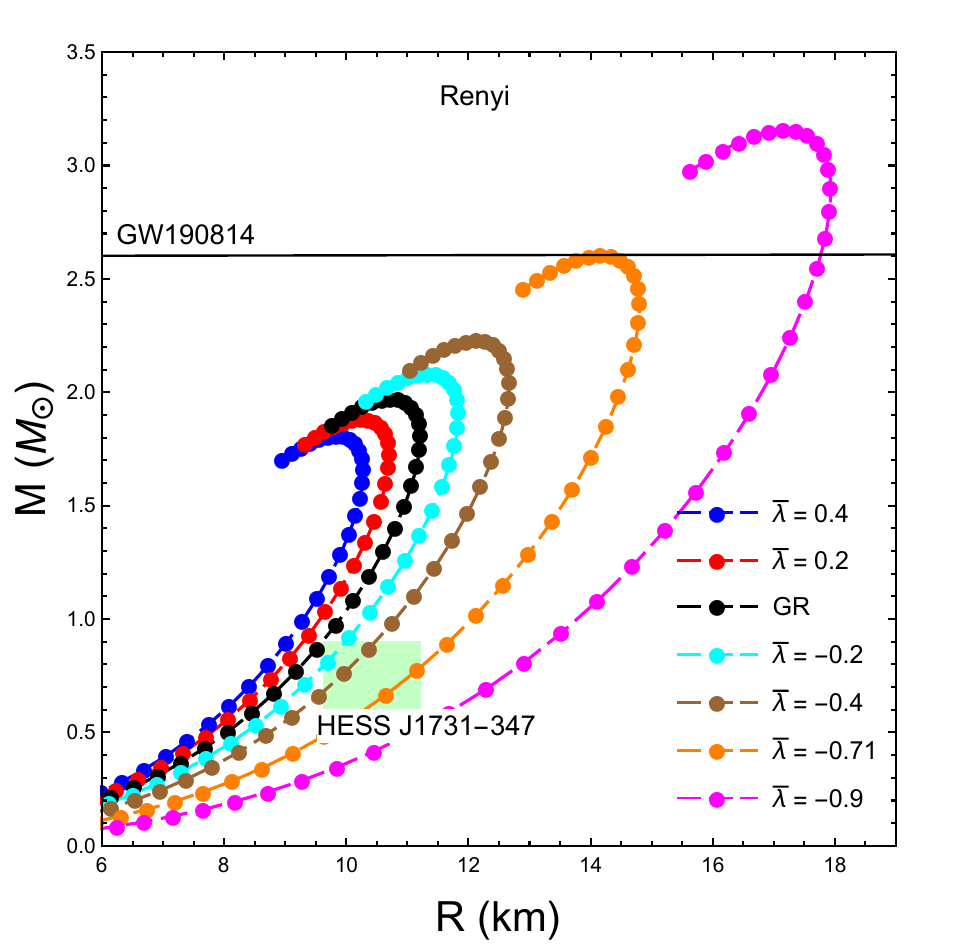}
	\includegraphics[width=0.45\textwidth]{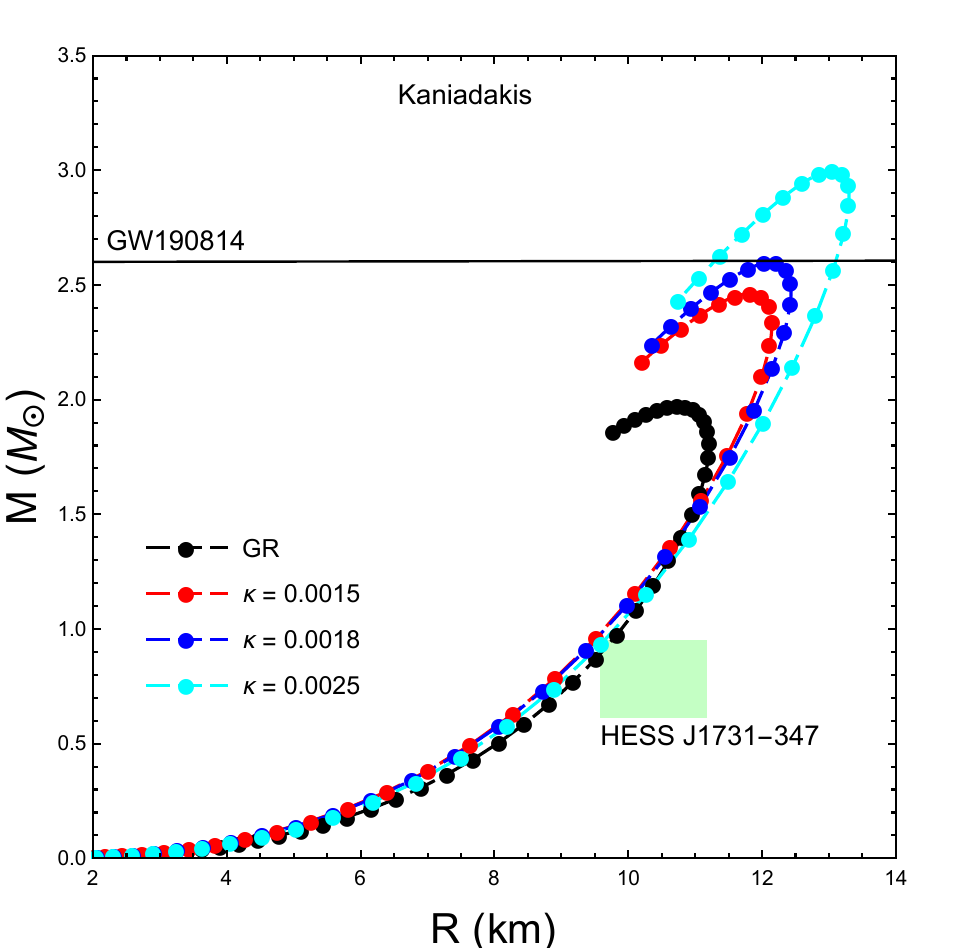}
	\caption{The mass-radius relation for QS in the $f(R)$ gravity model with the observable constraint from GW190814 and HESS J1731-347. Compared to GR (black dotted) for each model, the minimal length is adequate to lie within the upper and lower bounds. Notice that we use the Barrow and Tsallis parameter inside the range parameter shown in Section IV.}
    \label{fig:1}
\end{figure}

\begin{figure}[h!]
	\centering
	\includegraphics[width=0.48\textwidth]{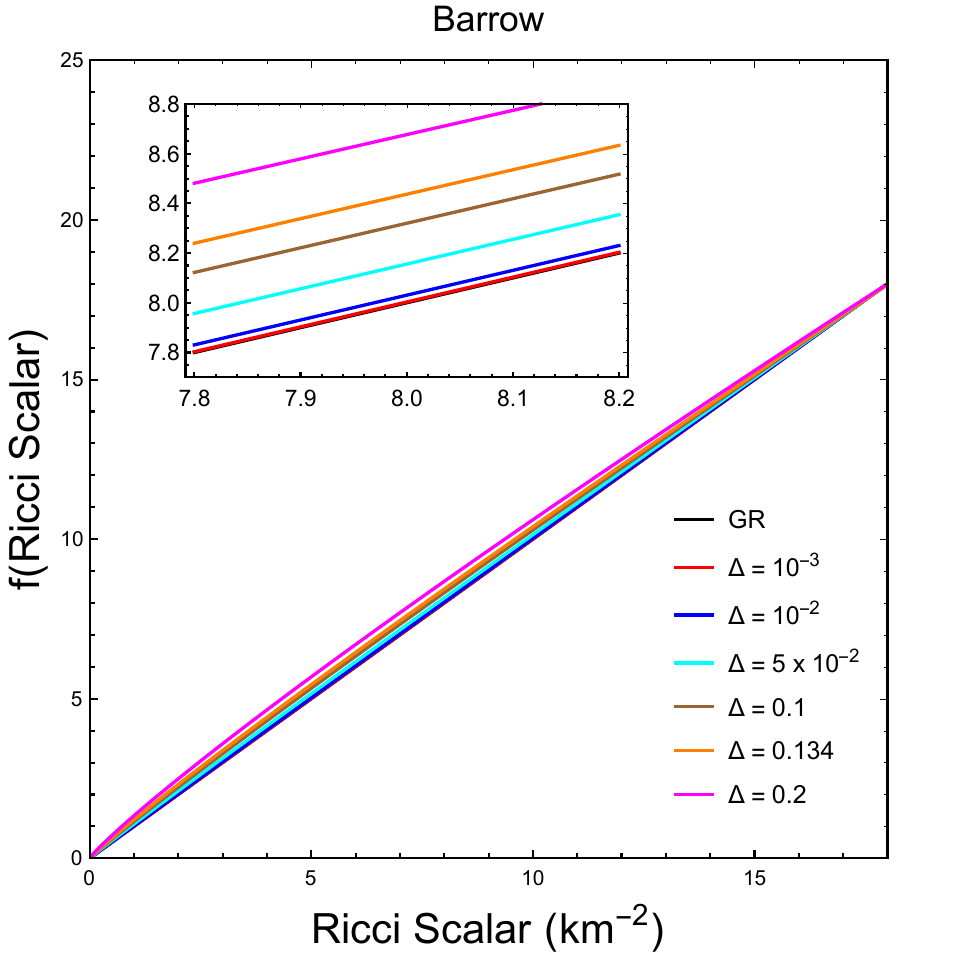}
	\includegraphics[width=0.48\textwidth]{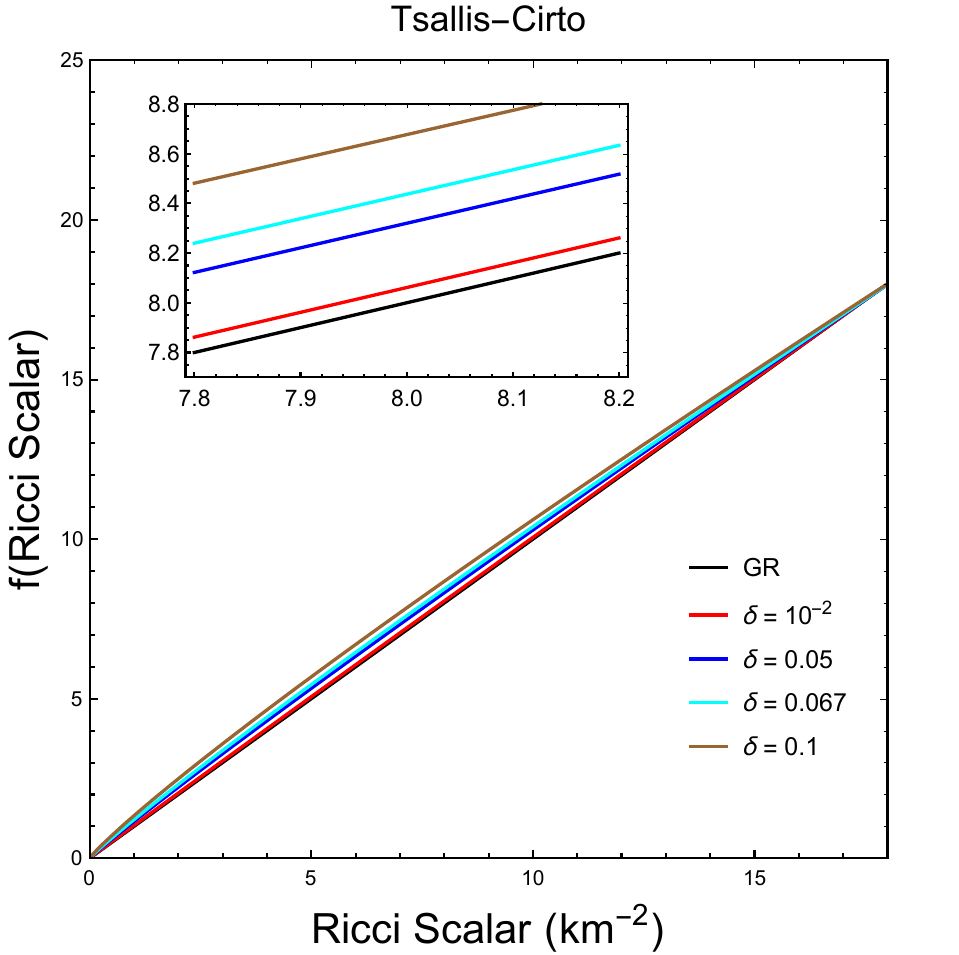}
	\includegraphics[width=0.48\textwidth]{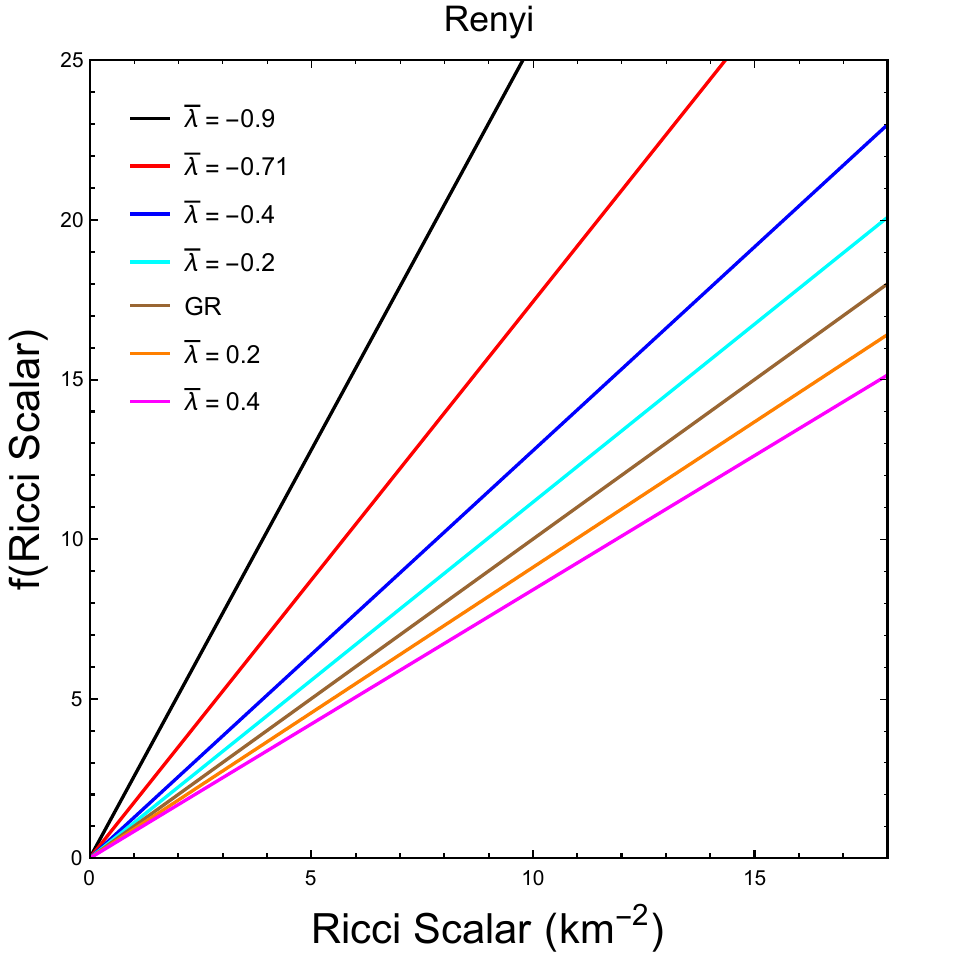}
	\includegraphics[width=0.48\textwidth]{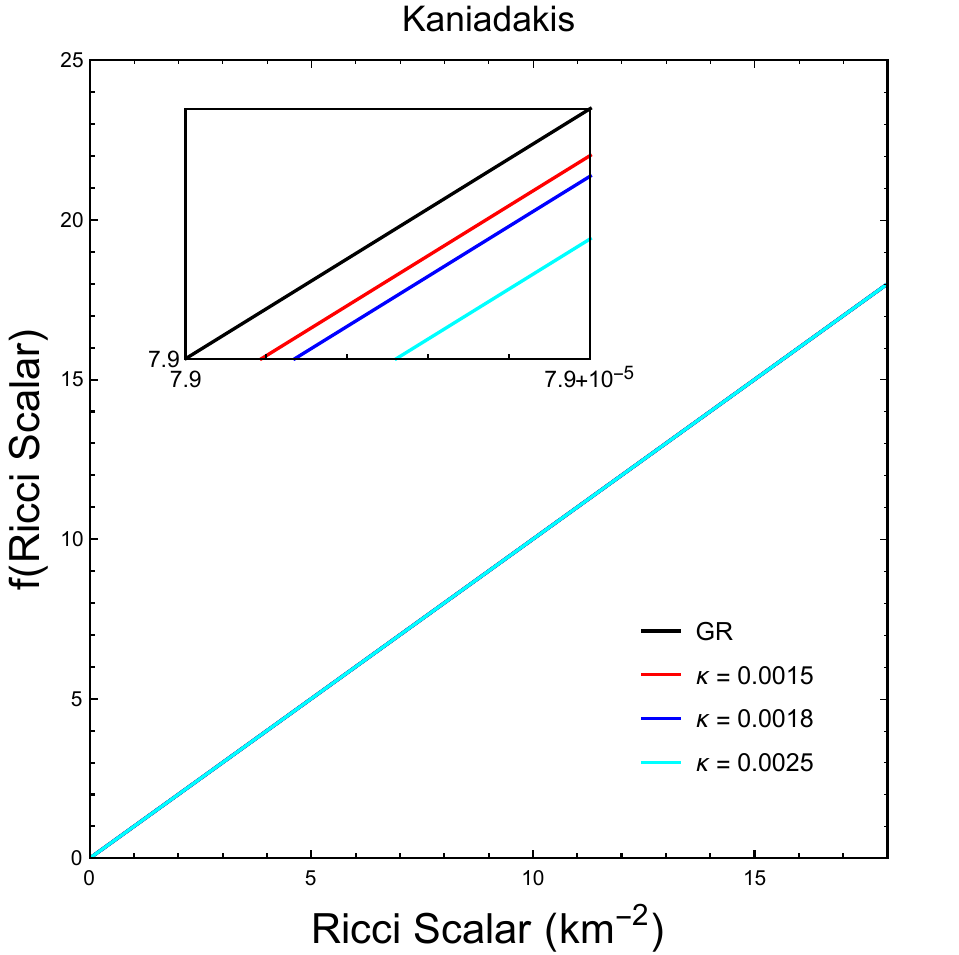}
	\caption{$f(R)$ vs $R$ relation for several representative entropy model, where $R$ in this context is the Ricci scalar. In the geometrized unit, the Ricci scalar $R$ has the unit $\textrm{km}^{-2}$.}
    \label{fig:1a}
\end{figure}

The EoS are now ready to be solved by integrating the field equations with the boundary conditions from the previous section, until we reach very small pressure, $p(R)\approx 10^{-7} ~\textrm{km}^{-2}$. With some central pressure intervals, we can obtain the mass-radius profile of the QS, which is plotted in Fig.~\ref{fig:1}, where we used four representative models with their parameters, and show the lower and upper bounds coming from observation HESS J1731-347 \cite{Doroshenko:2022nwp,Horvath:2023uwl} and GW190814 \cite{LIGOScientific:2020zkf}. The small inclusion from the parameter shows a discrepancy compared to the GR case. As we increase the central pressure further, we can see that each profile can lie within the minimum mass observed by HESS J1731, and the maximum mass lies in GW190814. If we look at each model in detail, Barrow and the Tsallis-Cirto model show a similar trend, with the rate of change unaltered at low central pressure. As we increase the central pressure, both profiles in each parameter deviate and increase until they reach the maximum mass. The Rényi entropy exhibits the same story, yet this model already deviates from the low central pressure. 
Lastly, the Kaniadakis entropy model shows a sensitive behavior, as it can increase the maximum mass with a small deviation in $\kappa$. All of the models exhibit distinct features due to the formalism of the $f(R)$ gravity model. Therefore, we try to analyze further by investigating the $f(R)$ vs $R$ relation in Fig.~\ref{fig:1a}, where $R$ in this context refers to the Ricci scalar quantity. We vary with the same parameter in Fig.~\ref{fig:1} to make it consistent with the analysis. The GR profile is recovered when we switch off the entropy parameter. For a nonzero entropy parameter, the behavior of the Barrow and Tsallis-Cirto models is quite similar, whereas the Rényi model changes drastically. It is mathematically expected, given the similarity with the GR case. In the Kaniadakis case, the deviation appears in a very small region. On the other hand, the exact values of each model within the GW190814 bound are reported in Table~\ref {tab:1}. We can see that the Kaniadakis model has the highest compactness within the bounds.

\begin{table}[h!]
	\centering
	\begin{tabular}{|c|c|c|c|c|c|}
		\hline
		
		\multirow{2}{*}{Model} 
		& \multicolumn{5}{c|}{Parameter} \\ \cline{2-6}
		
		& $\Delta,~\delta$,~$\bar{\lambda}$, and $\kappa$  
		& $p_c~(\textrm{km}^{-2})$ 
		& $M~(M_{\odot})$ 
		& $r_s $ (km) 
		& $\mathcal{C}$ \\
		
		\hline
		
		Barrow & $0.134$ & $3.07 \times10^{-4}$ & 2.61 & 13.72  & 0.281  \\ \hline
		Tsallis-Cirto & $0.067$ & $3.04 \times10^{-4}$ & 2.62 & 13.72 & 0.28 \\ \hline
		Renyi & $-0.712$ & $4 \times10^{-4}$ & 2.6 & 14.20 & 0.27 \\ \hline
		Kaniadakis & $0.0018$ & $2.2 \times10^{-4}$ & 2.6 & 12.12 & 0.32 \\ \hline
		
	\end{tabular}
	\caption{The numerical values at the GW190814 line. The parameters of Barrow, Tsallis-Cirto, Renyi, and Kaniadakis are denoted by the $\Delta,~\delta,~\bar{\lambda},$ and $\kappa,$ respectively.}
	\label{tab:1}
\end{table}
\begin{figure}[h!]
	\centering
	\includegraphics[width=0.48\textwidth]{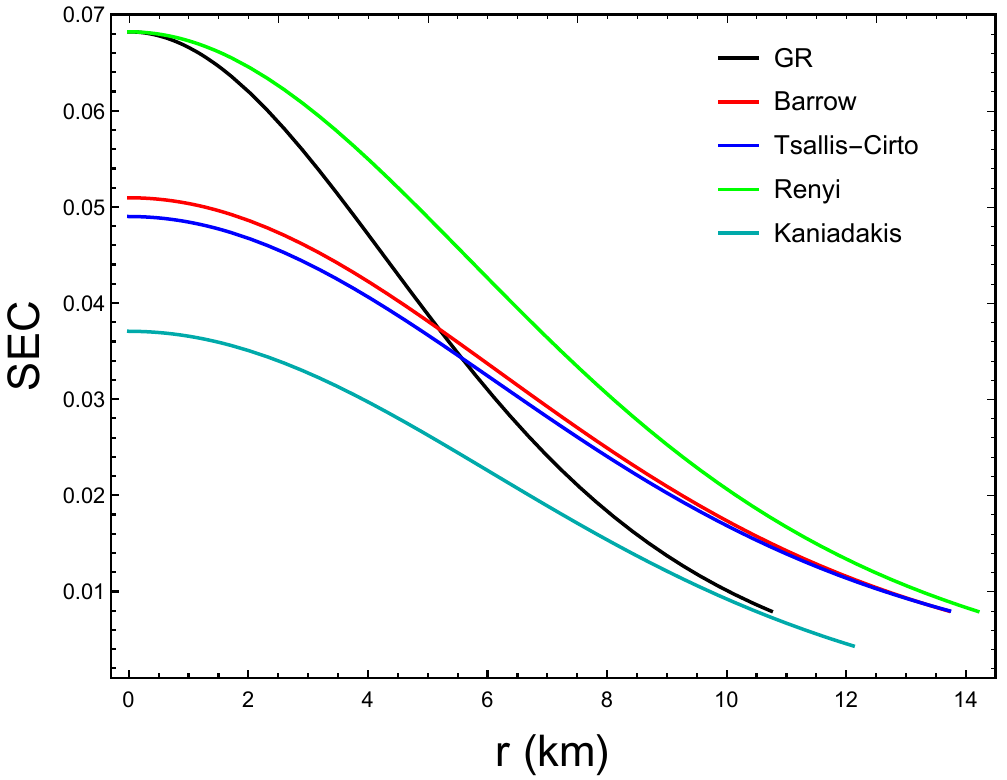}
	\includegraphics[width=0.48\textwidth]{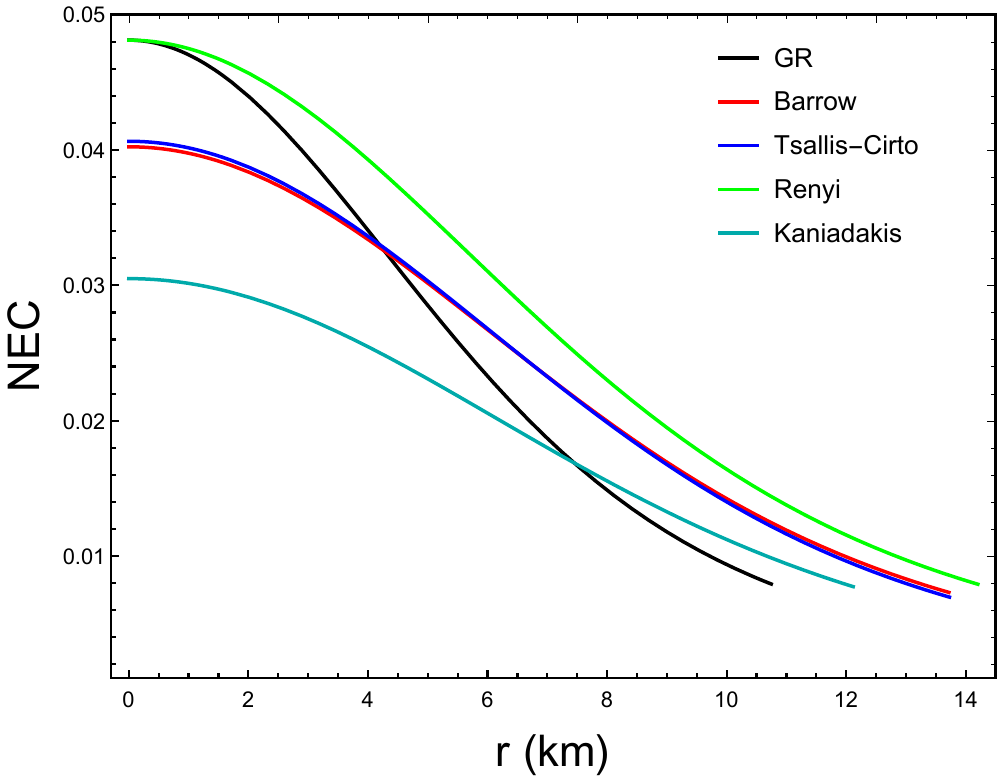}
	\includegraphics[width=0.48\textwidth]{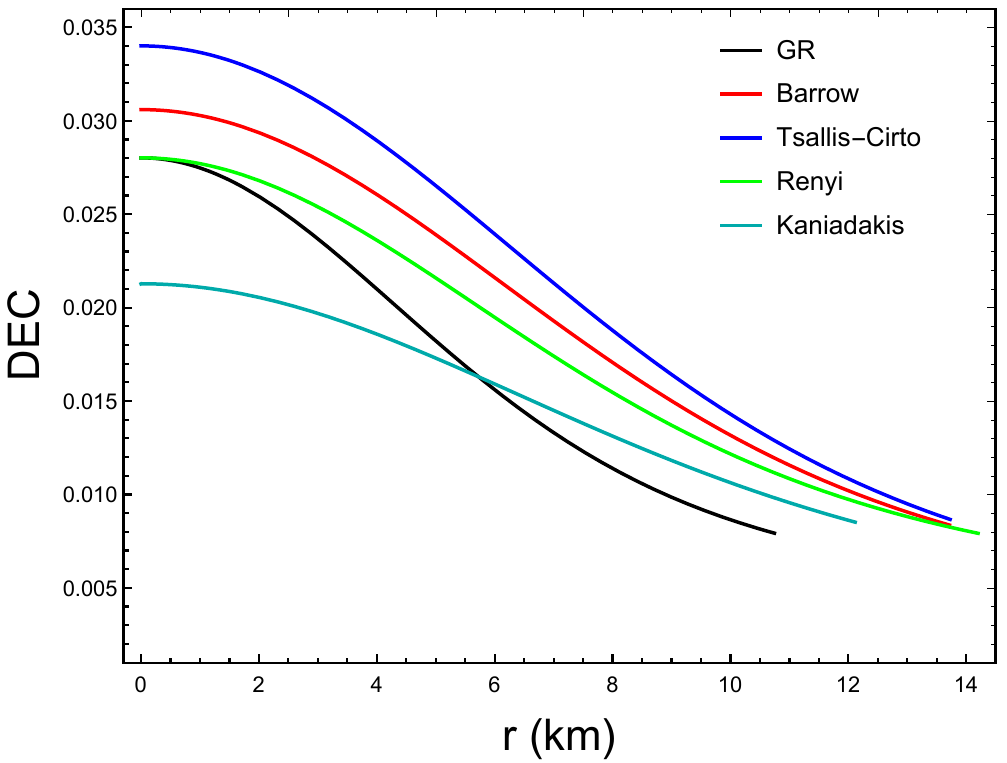}
	\includegraphics[width=0.48\textwidth]{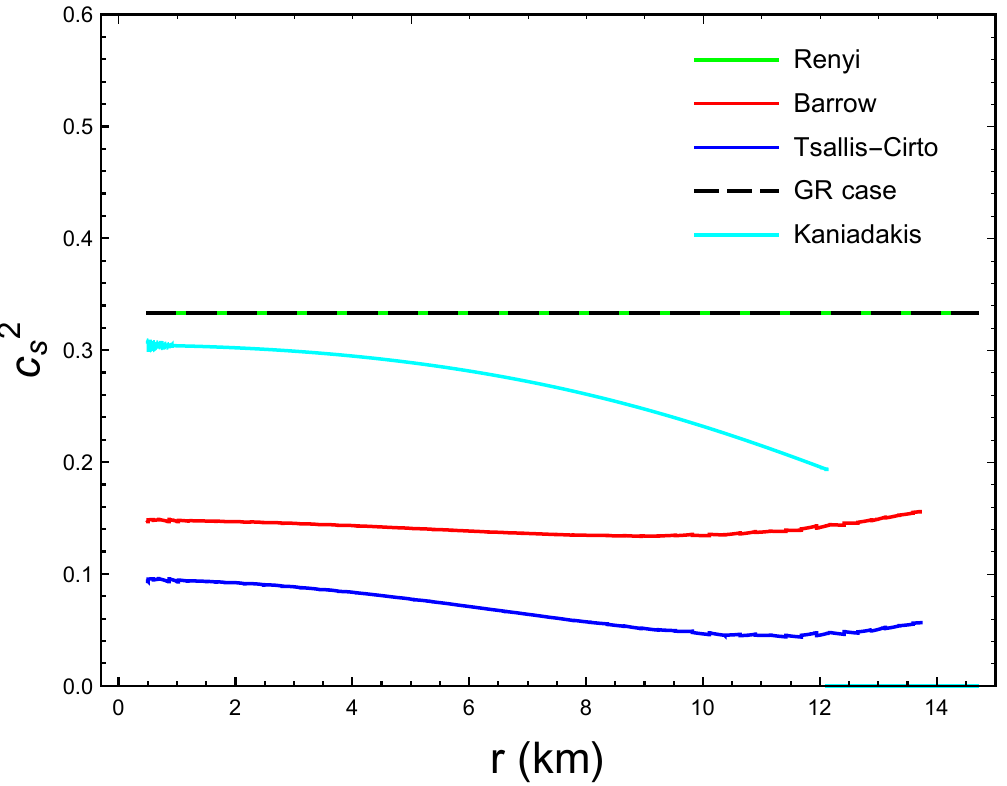}
	\caption{The energy conditions of four representative minimal length model. Each of the profile plotted here is corresponded to the maximum mass that lie within both bound (GW190814 and HESS J1731-347). In the bottom right, we show the speed of sound with the same parameter. It is shown that, at the maximum mass, the profiles are satisfied energy conditions and the speed of sound is below unity.}
    \label{fig:2EC}
\end{figure}

To ensure realistic conditions, we also compute the energy conditions for generalized $f(R)$ (see \cite{Santos:2007bs} for a review). Unlike the GR case, the general energy conditions are embedded in the effective tensor energy-momentum. In the $f(R)$ gravity model, the field equations can be recast into effective equations
\begin{equation}
    R_{\mu\nu} = \left( \mathcal{T}_{\mu\nu} - \frac{1}{2}g_{\mu\nu}\mathcal{T}\right),
\end{equation}
where
\begin{eqnarray}
    \mathcal{T} &=&\frac{1}{8\pi  f_{R}}(8\pi T+f-Rf_{R}-3\Box f_{R}),\\ \mathcal{T}_{\mu\nu} &=& \frac{1}{8\pi f_{R}}[8\pi T_{\mu\nu}+(\nabla_{\mu}\nabla_{\nu}-g_{\mu\nu}\Box)f_{R}]. 
\end{eqnarray}
Therefore, the strong energy condition (SEC), the null energy condition (NEC), and the dominant energy condition (DEC) can be written as 
\begin{equation}
	8\pi(\rho+3p) + f-R f_{R} - \nu'e^{-\lambda} R' f_{RR} - e^{-\lambda} \bigg[ \left( \frac{2}{r} + \frac{\nu'-\lambda'}{2} \right)R' f_{RR} + R'' f_{RR} + R'^2 f_{RRR}   \bigg] \geq0.
\end{equation}
\begin{equation}
	8\pi (\rho+p) + e^{-\lambda} \bigg[ R'' f_{RR} + R'^2 f_{RRR} - \frac{(\nu'+\lambda')}{2}R' f_{RR}  \bigg] \geq0,
\end{equation}
\begin{equation}
	8\pi(\rho-p) -f+R f_{R} + e^{-\lambda} \bigg\lbrace \left[ \frac{4}{r} - \frac{(\lambda'-\nu')}{2} \right] R' f_{RR} + R''f_{RR} +R'^2 f_{RRR}  \bigg\rbrace \geq0.
\end{equation}

\begin{figure}[h!]
	\centering
    \includegraphics[width=0.48\textwidth]{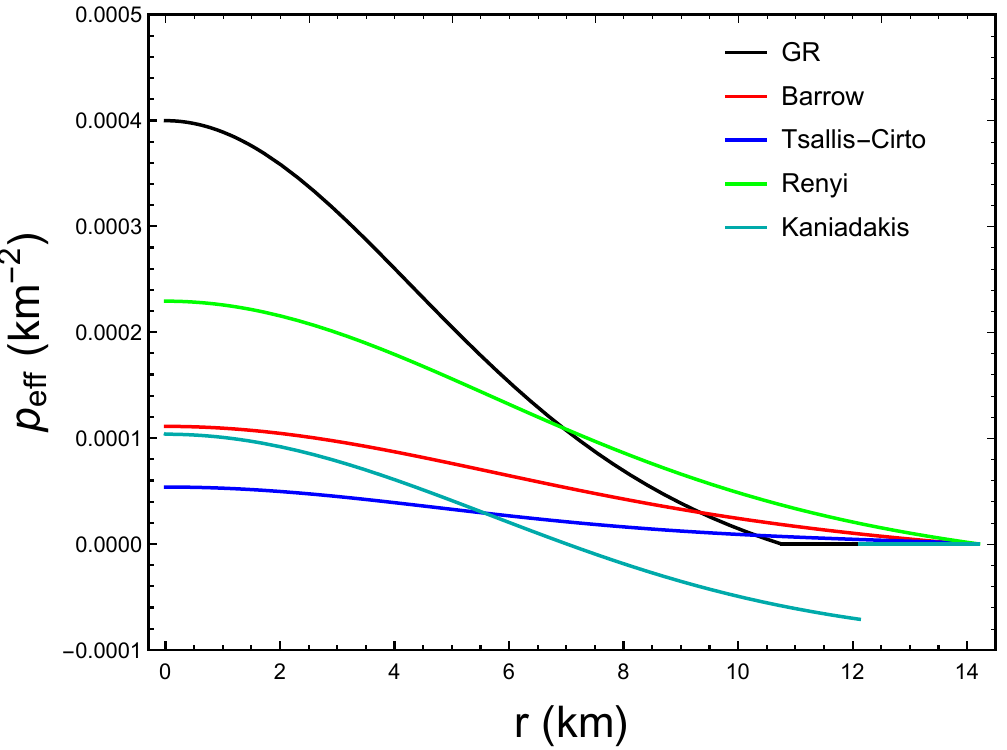}
    \includegraphics[width=0.48\textwidth]{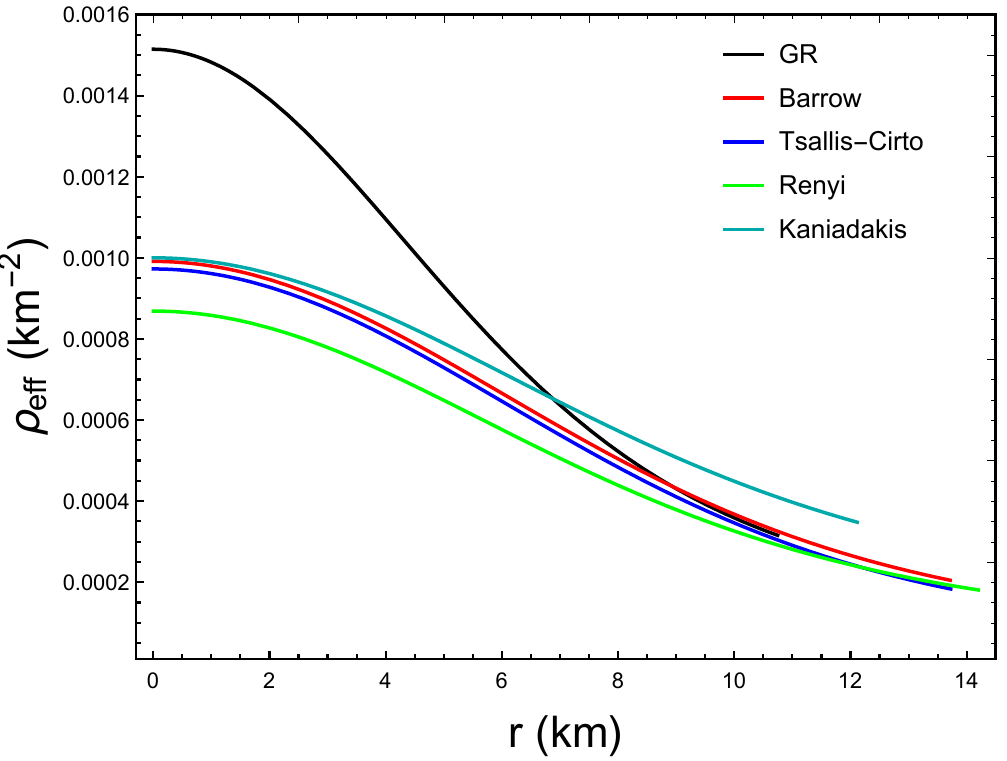}
    \includegraphics[width=0.48\textwidth]{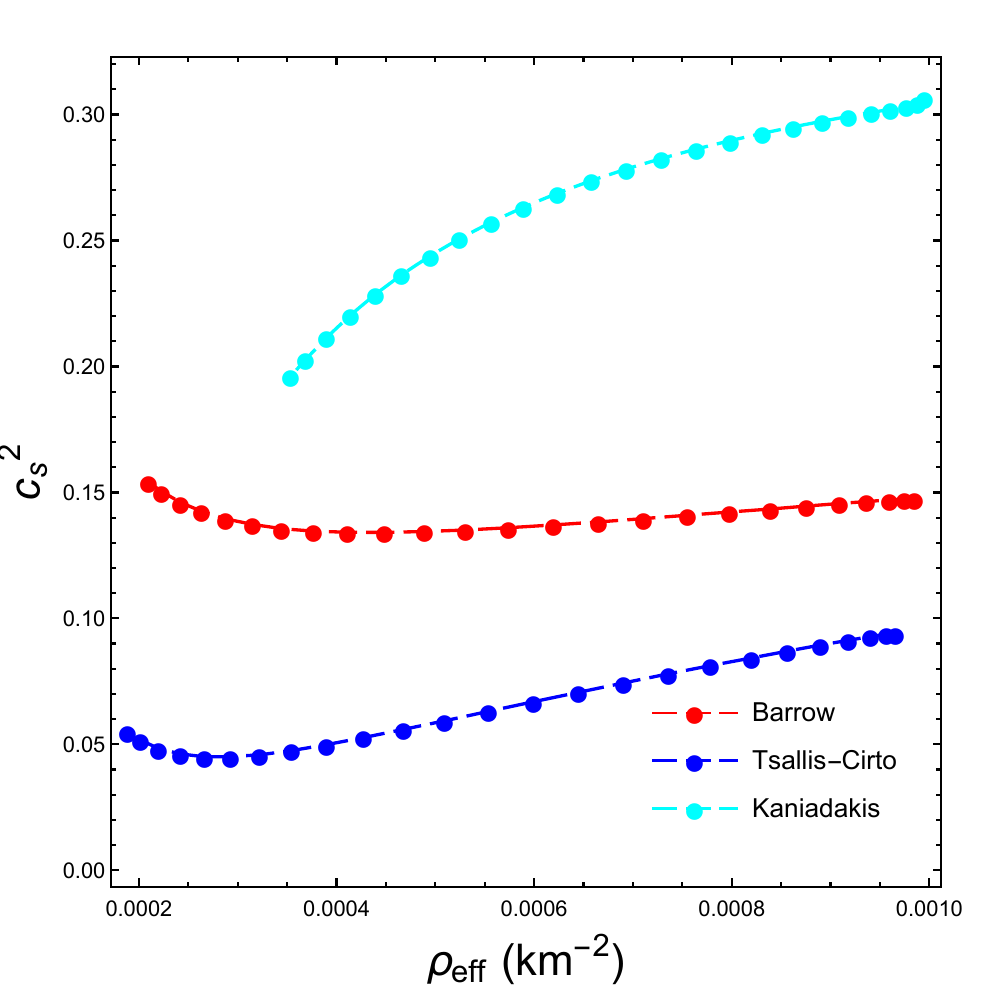}
	\caption{Extension analysis of the speed of sound presented in Fig.~\ref{fig:2EC}, where we consider the effective profiles versus radial coordinate and the speed of sound versus the effective density profile.}
    \label{fig:3}
\end{figure}

The energy conditions for GR \cite{Hawking:1973uf} are recovered when we choose the generalized model to be $f(R)=R$\footnote{Note that, to make it consistent, we include the factor $8\pi$ in the GR energy conditions.} From these equations, the general $f(R)$ model employs a formalism distinct from GR, involving the $f-$ term, the metric's derivative, and its derivative with respect to the Ricci scalar. Profile illustrations are shown in Fig.~\ref{fig:2EC}. The investigated parameter is similar to Table~\ref{tab:1}; each profile is plotted over its radial range, $r \in [0, r_s]$. When these profiles satisfy the conditions, the corresponding central pressures are usually automatically satisfied as well. All considered models satisfy the energy conditions. Another realistic test is the speed of sound: the latter quantity, $c_{s}^2 \equiv dp_{\textrm{eff}}/d\rho_{\textrm{eff}} \leq 1$, where
\begin{eqnarray}
  \rho_{\textrm{eff}} &=& \frac{1}{8\pi f_R}\bigg\lbrace 8\pi\rho-\frac{1}{2}(f-Rf_{R}) + e^{-\lambda} \bigg[\left(\frac{2}{r}-\frac{\nu'}{2} \right) \bigg]R'f_{RR} + R'' f_{RR}+R'^2 f_{RRR}  \bigg\rbrace, \\ \textrm{and},\\
  p_{\textrm{eff}} &=& \frac{1}{8\pi f_R}\bigg\lbrace 8\pi p+\frac{1}{2}(f-Rf_{R}) - e^{-\lambda} \bigg[\left(\frac{2}{r}+\frac{\nu'}{2} \right) \bigg]R'f_{RR} + R'' f_{RR}+R'^2 f_{RRR}  \bigg\rbrace, 
\end{eqnarray}
is useful to ensure consistency with subluminal conditions. It is also worth noting that the Rényi entropy matches the GR counterpart, because the chosen $f(R)$ function is simply proportional to the Ricci scalar. In contrast, distinct properties arise in the Barrow, Tsallis-Cirto, and Kaniadakis models due to their unique entropy formulations. The lowest profile of the speed of sound, indicated by the blue solid line, corresponds to the model with the smallest central density among the three, as shown in Fig.~\ref{fig:3} for $c_s^2$ vs $\rho_{\textrm{eff}}$. Extension analysis can be performed by examining the effective profiles as functions of the radial coordinate in Fig.~\ref{fig:3}. The effective pressure becomes zero at the surface. We observe that all effective densities remain positive; however, Kaniadakis's effective pressure is negative for around half its values. Therefore, a physically realistic compact object can only occur when the quark star is described by the Barrow, Tsallis-Cirto, or Rényi entropy models, each of which shows distinct behavior in its effective pressure and density.

\section{Conclusion}

We established a direct relationship between horizon entropy and $f(R)$ gravity for four representative non-extensive entropy models. The generic structure of $f(R)$ demonstrates distinctive mathematical characteristics, as observed in the mass-radius (M-R) diagram and the $f(R)$ versus $R$ plot for the quark star (QS) model. The non-extensive entropy falls within the lower limit from HESS J1731-347 and the upper threshold from the GW190814 gravitational-wave event, while general relativity (GR) is only marginally within the lower constraint. We subsequently evaluate the classical energy conditions and causality (subluminal sound speed) to validate that the parameter profiles inside the bounds are physically consistent. Results indicate all profiles fulfill the energy conditions. For GR, the speed of sound for the MIT bag model quark star remains constant, as does that of the Rényi model. This mathematical behavior arises from the proportional relationship of the chosen $f(R)$ function to the Ricci scalar. Therefore, quark stars characterized by Barrow, Tsallis-Cirto, and Rényi entropies constitute theoretical models that could describe exotic compact objects such as HESS J1731-347 (low-mass) and the secondary body in GW190814 (heavy-mass), respectively.

\acknowledgments
We thank Muhammad Fahmi Fauzi and Faris Ramadhantyo Darmawan for the useful discussions. In this work, we are supported by Hibah Fundamental DIKTI, PKS-178/UN2.RST/HKP.05.00/2026. \\

\end{document}